\documentclass[twocolumn,showpacs,prc,superscriptaddress,amsmath,amssymb]{revtex4-1}

\usepackage{multirow}
\usepackage{graphicx}
\usepackage{dcolumn}
\usepackage{bm}
\usepackage{soul}
\usepackage{color}

\begin{document}

\preprint{APS/123-QED}

\title{Exploiting neutron-rich radioactive ion beams to constrain the symmetry energy}

\author{Z. Kohley}
 \email{kohley@nscl.msu.edu}
\affiliation{National Superconducting Cyclotron Laboratory, Michigan State University, East Lansing, Michigan 48824, USA}
\affiliation{Department of Chemistry, Michigan State University, East Lansing, Michigan 48824, USA}
\author{G.~Christian}
    \altaffiliation[Present address: ]{TRIUMF, 4004 Wesbrook Mall, Vancouver, British Columbia V6T 2A3, Canada}
	\affiliation{National Superconducting Cyclotron Laboratory, Michigan State University, East Lansing, Michigan 48824, USA}
	\affiliation{Department of Physics \& Astronomy, Michigan State University, East Lansing, Michigan 48824, USA}
\author{T.~Baumann}
	\affiliation{National Superconducting Cyclotron Laboratory, Michigan State University, East Lansing, Michigan 48824, USA}
\author{P.A.~DeYoung}
	\affiliation{Department of Physics, Hope College, Holland, Michigan 49423, USA}
\author{J.E.~Finck}
    \affiliation{Department of Physics, Central Michigan University, Mt. Pleasant, Michigan, 48859, USA}
\author{N.~Frank}
	\affiliation{Department of Physics \& Astronomy, Augustana College, Rock Island, Illinois, 61201, USA}
\author{M.~Jones}
	\affiliation{National Superconducting Cyclotron Laboratory, Michigan State University, East Lansing, Michigan 48824, USA}
	\affiliation{Department of Physics \& Astronomy, Michigan State University, East Lansing, Michigan 48824, USA}
\author{J.~K.~Smith}
	\affiliation{National Superconducting Cyclotron Laboratory, Michigan State University, East Lansing, Michigan 48824, USA}
	\affiliation{Department of Physics \& Astronomy, Michigan State University, East Lansing, Michigan 48824, USA}
\author{J.~Snyder}
	\affiliation{National Superconducting Cyclotron Laboratory, Michigan State University, East Lansing, Michigan 48824, USA}
	\affiliation{Department of Physics \& Astronomy, Michigan State University, East Lansing, Michigan 48824, USA}
\author{A.~Spyrou}
	\affiliation{National Superconducting Cyclotron Laboratory, Michigan State University, East Lansing, Michigan 48824, USA}
	\affiliation{Department of Physics \& Astronomy, Michigan State University, East Lansing, Michigan 48824, USA}
\author{M.~Thoennessen}
	\affiliation{National Superconducting Cyclotron Laboratory, Michigan State University, East Lansing, Michigan 48824, USA}
	\affiliation{Department of Physics \& Astronomy, Michigan State University, East Lansing, Michigan 48824, USA}

\date{\today}

\begin{abstract}
 The Modular Neutron Array (MoNA) and 4~Tm Sweeper magnet were used to measure the free neutrons and heavy charged particles from the radioactive ion beam induced $^{32}$Mg~+~$^{9}$Be reaction.  The fragmentation reaction was simulated with the Constrained Molecular Dynamics model (CoMD), which demonstrated that the $\langle N/Z \rangle$ of the heavy fragments and free neutron multiplicities were observables sensitive to the density dependence of the symmetry energy at subsaturation densities.  Through comparison of these simulations with the experimental data constraints on the density dependence of the symmetry energy were extracted.  The advantage of radioactive ion beams as a probe of the symmetry energy is demonstrated through examination of CoMD calculations for stable and radioactive beam induced reactions.
\end{abstract}


\maketitle
\par
\emph{Introduction.} The desire to extend our understanding of nuclear matter at densities, temperatures, pressures, and neutron-to-proton ratios ($N/Z$) away from that of ground state nuclei has become a driving force of the nuclear science community.  In particular, the emergence of radioactive ion beam (RIB) facilities has placed an emphasis on exploring nuclear matter along the isospin degree-of-freedom.  The symmetry energy is the critical component which defines how the properties of nuclear matter, or the nuclear Equation of State (EoS), change as a function of isospin. The nuclear EoS can be approximated as
\begin{equation}
 E(\rho,\delta) = E(\rho,0) + E_{sym}(\rho)\delta^{2} + O(\delta^{4})
\end{equation}
where the energy per nucleon of infinite nuclear matter, $E(\rho,\delta)$, is a function of the density ($\rho$) and isospin concentration ($\delta$)~\cite{LI01,LI08,FUCH06}. The isospin concentration is the difference in the neutron and proton densities, $\delta = (\rho_{n} - \rho_{p}) / (\rho_{n} + \rho_{p}) \approx (N-Z)/A$.  The first term of the EoS is isopin-independent and thus represents the binding energy of symmetric ($N=Z$) nuclear matter.  The second term of the EoS has a strong dependence on the asymmetry of the nuclear matter and has been historically termed the symmetry energy (however, a more appropriate term is the ``asymmetry energy'').  The EoS for symmetric nuclear matter is relatively well constrained around the saturation density ($\rho_{0} = 0.16$~fm$^{-3}$) from isoscalar giant monopole and dipole resonances~\cite{SHLOMO06,Khan12} and at higher densities (up to $\rho / \rho_{0} \sim$4.5) from heavy-ion collisions~\cite{DANI02,Har06}.

\par
Constraining the form of the density dependence of the symmetry energy ($E_{sym}(\rho)$) is essential for developing a complete description of asymmetric nuclear matter.  For example, the properties of neutron stars are strongly correlated to the density dependence of the symmetry energy~\cite{Lat07,Latt12,Stei12}.  Theoretical calculations have shown that the cooling process~\cite{LI0222,Rob12}, the mass-radius relationship~\cite{Latt01,LI06}, outer crust composition~\cite{Roca08}, and core-crust transition properties~\cite{Duc11} are all linked to the form of $E_{sym}(\rho_{0})$.  Similarly, properties of neutron-rich (or proton-rich) nuclei, such as the neutron skin thickness, are dependent on the density dependence of the symmetry energy~\cite{Horo01,Chen10,Roc11}.

\par
While a number of different observables have been used to explore the symmetry energy (see the review in Ref.~\cite{Tsang12}), heavy-ion collisions are particularly important due to their ability to probe different density regions of nuclear matter~\cite{LI08,Toro10}.  Significant progress was made by Tsang \emph{et al.} using neutron-to-proton ratios and isospin diffusion observables from heavy-ion collisions to obtain constraints on the density dependence of the symmetry energy at subsaturation densities~\cite{TSANG09,Tsang12}.  While the use of heavy-ion collisions as a probe of the symmetry energy has been demonstrated numerous times~\cite{SHETTY07,FAMIANO06,Amo09,KOHLEYIMF,KOHLEYLCP,RUSSOTTO11}, experiments using radioactive ion beam induced reactions have remained relatively unexplored.  In comparison, theoretical works have commonly used radioactive ion beam induced reactions to demonstrate their sensitivity to the density dependence of the symmetry energy~\cite{Li97,Col98,LI022,Gre03,Chen03,Chen032,Ono042}.  For example, a $^{132}$Sn~+~$^{132}$Sn reaction will exhibit a much larger sensitivity to the symmetry energy in comparison to a $^{124}$Sn~+~$^{124}$Sn reaction due to the increased $\delta$ of the $^{132}$Sn system.  The work of the ALADIN Collaboration represents the only experimental example in which RIB induced reactions ($^{124}$La and $^{107}$Sn) were used to examine the symmetry energy~\cite{Sfi09,Ogul11,Traut11}.  The experimental results were shown to be sensitive to the strength of the symmetry energy through comparisons with the statistical multifragmentation model~\cite{Ogul11,Traut11} and the isospin quantum molecular dynamics model~\cite{Kumar12}.

\begin{figure*}
\includegraphics[width=0.99\textwidth]{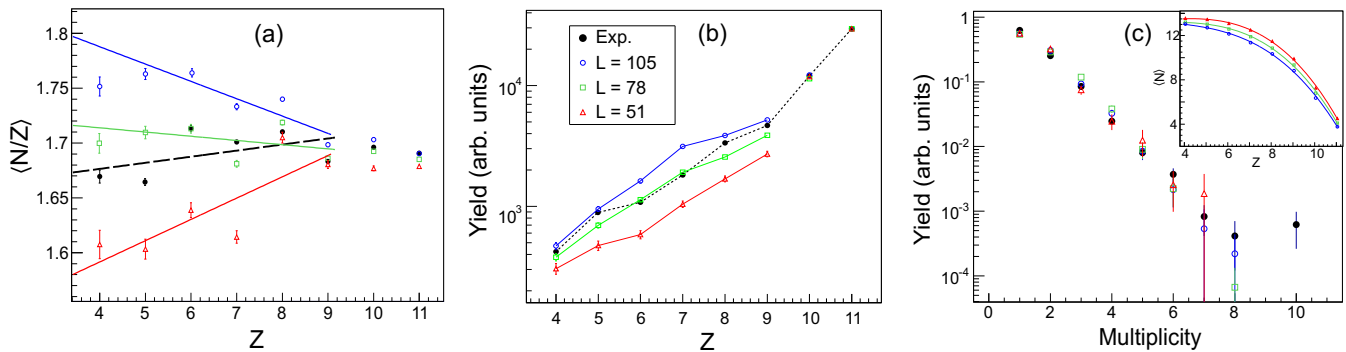}
\caption{\label{f:exp} (Color online) (a) Average $N/Z$ as a function of the charge of the fragments. Linear fits to the experimental (dashed black line) and CoMD (solid colored lines) results are shown to guide the eye. (b) Yield of the different elements produced in the $^{32}$Mg~+~$^{9}$Be reaction normalized to the $Z = 11$ fragments.  (c) Hit multiplicity distribution measured in MoNA in coincidence with the $Z = 9$ fragments normalized to the multiplicity~=~1 events.  In each panel, the experimental results are shown as solid black circles and the open data symbols represent the filtered CoMD simulations with different values of $L$, as described by the legend.  In panels (a) and (b), lines are only shown for the $Z = 4 -9$ fragments denoting the region used to extract constraints on $E_{sym}(\rho)$.  In the insert of panel (c) the unfiltered average neutron multiplicity is shown as a function of the charge of the residue from the CoMD simulation. }
\end{figure*}

\par
In this article, constraints on the density dependence of the symmetry energy are extracted from the $^{32}$Mg~+~$^{9}$Be RIB induced fragmentation reaction through a comparison with the Constrained Molecular Dynamics (CoMD) model~\cite{PAPA01,PAPA05}.  The limits extracted from the present work provide the first confirmation of the heavy-ion constraints put forth by Tsang~\textit{et al.}~\cite{TSANG09}.  The $^{24}$Mg~+~$^{9}$Be stable beam reaction was also simulated and compared to the $^{32}$Mg reaction to demonstrate the advantage RIBs offer as a probe of the symmetry energy.

\par
\emph{Experiment.} The radioactive $^{32}$Mg beam was produced by the Coupled Cyclotron Facility at the National Superconducting Cyclotron Laboratory at Michigan State University.  A 140 MeV/nucleon $^{48}$Ca primary beam bombarded a 1316 mg/cm$^{2}$ Be production target producing a wide range of fragmentation products.  The A1900 fragment separator allowed the secondary beam of interest, $^{32}$Mg, to be selected from the fragmentation products and unreacted primary beam.  The 73 MeV/nucleon $^{32}$Mg beam bombarded a 288 mg/cm$^{2}$ $^{9}$Be secondary target.  The charged particle residues and free neutrons resulting from the fragmentation reaction were measured in a suite of charged particle detectors placed downstream of the 4~Tm Sweeper magnet~\cite{SWEEPER} and in the Modular Neutron Array (MoNA)~\cite{MONA03,MONA05}, respectively.   A detailed description of the experimental setup, calibrations, and particle identification procedure are presented in Ref.~\cite{Chri12}.

\par
\emph{Results and Discussion.} In the first step of the fragmentation process, nucleons will be removed (or picked-up) from the $^{32}$Mg projectiles through collisions with the $^{9}$Be nuclei producing excited projectile-like fragments.  These excited fragments will then decay, through the emission of light charged particles, neutrons and $\gamma$-rays, leaving a projectile-like residue. The final $N/Z$ of the residues and the related number of neutrons emitted should be sensitive to the density dependence of the symmetry energy.  Assuming that the hot projectile-like fragments expand to a density below $\rho_{0}$, a soft density dependence of the symmetry energy will be more repulsive leading to an increase in neutron emission and, therefore, a decreased $N/Z$ of the residues.  A stiff density dependence of the symmetry energy will be less repulsive resulting in a decreased neutron emission and increased $N/Z$ of the residues.  A limitation of the CoMD approach is that it does not explicitly treat the direct population of unbound resonant states, however examination of the experimental data showed the neutron decay spectra to be dominated by continuum or thermal distributions.
\par
In Fig.~\ref{f:exp}(a) the average $N/Z$ of the detected residues is shown as a function of their charge from the experimental data.  It should be recognized that only a selection of the mass distribution for each $Z$ was measured based on the roughly $\pm$8$\%$ $B\rho$ (magnetic rigidity) acceptance of the Sweeper magnet.  Thus, the $\langle N/Z \rangle$ was constructed from the three to four isotopes for each $Z$ that were within the $B\rho$ acceptance.  In Fig.~\ref{f:exp}(b) the total yield of each element is presented and in Fig.~\ref{f:exp}(c) the multiplicity distribution measured in MoNA in coincidence with the $Z = 10$ fragments is shown.

\par
The $^{32}$Mg~+~$^{9}$Be reaction was simulated using the CoMD model in which each nucleon is represented by a Gaussian wave-packet that is propagated following the equations of motion derived from the time-dependent varational principle~\cite{PAPA01,PAPA05}.  Each event was simulated for 1500~fm/c allowing for the excited projectile-like fragments to be produced and then decay within one consistent description of the reaction.  These fragmentation type reactions have been previously shown to be well reproduced by molecular dynamics simulations~\cite{MOCKO08}.  At lower energies (15$-$25~MeV/nucleon), Souliotis \textit{et al.} has shown the CoMD simulation to accurately reproduce the properties of residues produced from binary deep-inelastic transfer reactions~\cite{Soul11,Soul10}.

\par
For comparison with the experimental data, the results from the CoMD simulation were filtered through a software replica of the experimental setup which took into account the $B\rho$ and angular acceptances of the charged particles.  The $B\rho$ and angular acceptance of the measured fragments were determined using inverse tracking maps produced from the Cosy Infinity ion-optics program~\cite{Frank07}.  Additionally, the efficiencies of the charged particle detectors as a function of $Z$ were determined and applied to the simulation.   A $\texttt{\sc Geant4}$ simulation was used to simulate the interactions of the neutrons in MoNA.  This allows for multiple neutron elastic and inelastic scattering reactions in MoNA to be reproduced correctly.  Additional details and validation of the $\texttt{\sc Geant4}$ simulation for neutron interactions in MoNA can be found in Refs.~\cite{Koh12,Koh13}.

\par
The CoMD model results are compared to the experimental data in Fig.~\ref{f:exp}(a)-(c).  In the CoMD model three different forms of isospin-dependent part of the effective nucleon-nucleon interaction can be selected and are labeled by the the associated slope of the symmetry energy at saturation density defined as $L = 3 \rho_{\circ} \frac{\partial E_{sym}(\rho)}{\partial \rho} \lvert _{\rho_{\circ}}$.  Additional details about the different forms of the symmetry energy, such as the curvature ($K_{sym}$), can be found in Ref.~\cite{Kohley12}. The magnitude of the symmetry energy at saturation density, $E_{sym}(\rho_{0})$, was set at 32~MeV.  Calculations using other values of $E_{sym}(\rho_{0})$ have been performed, and will be discussed subsequently.  As shown, the $\langle N/Z \rangle$ of the fragments produced in the RIB induced reaction are strongly dependent on the form of the symmetry energy.  The results follow the trend discussed above with the soft density dependence of the symmetry energy ($L = 51$~MeV) resulting in the lowest $\langle N/Z \rangle$.  This confirms the expectation that the neutron emission or $\langle N/Z \rangle$ of the residues probes the density dependence of the symmetry energy below saturation density.  The density region probed in the experiment was also estimated from the CoMD simulations to be slightly below saturation density ($\rho/\rho_{0} > 0.5$).  Thus, the describing of the form of the symmetry energy in terms of $L$ should be appropriate.

\par
The elemental yield of the fragments measured in the experiment (Fig.~\ref{f:exp}(b)) are also observed to be sensitive to the form of the symmetry energy used in the CoMD simulation. The sensitivity of the $Z$ yields is dominated by $B\rho$ acceptance which selects a portion of the isotopic distributions for each $Z$.  Thus, if the isotopic distribution is peaked within the acceptance of the experiment a larger fragment yield is observed (as is the case for $L = 105$~MeV).


\par
Following the sensitivity of the $\langle N/Z \rangle$ of the fragments, it would be expected that the associated neutron multiplicity is also sensitive to the form of the symmetry energy.  The experimental and filtered CoMD neutron multiplicity distributions in coincidence with $Z = 9$ residues are presented in Fig.~\ref{f:exp}(c).  While the distributions agree well, very little dependence on $L$ is observed from the CoMD simulation.  In the insert of Fig.~\ref{f:exp}(c) the unfiltered average neutron multiplicity as a function of the residue $Z$ from the CoMD simulations is shown and the expected dependence on $L$ is recovered.  The lower (higher) $L$ results in an increased (decreased) neutron emission which is correlated to the decreased (increased) $\langle N/Z \rangle$ of the fragments.  The lack of sensitivity of the filtered CoMD results to $L$ was examined and found to be related to the multiple scattering of neutrons in MoNA, the angular acceptance of MoNA, and the coincidence trigger used in the data collection, which removed all zero multiplicity events.  A future planned experiment will increase the sensitivity of the neutron measurements to the symmetry energy through use of a second neutron array providing increased angular acceptance.  Also a larger range of isotopes will be measured from multiple $B\rho$ settings allowing for neutron measurements in coincidence with more neutron-deficient fragments.

\par
\begin{figure}
\includegraphics[width=0.48\textwidth]{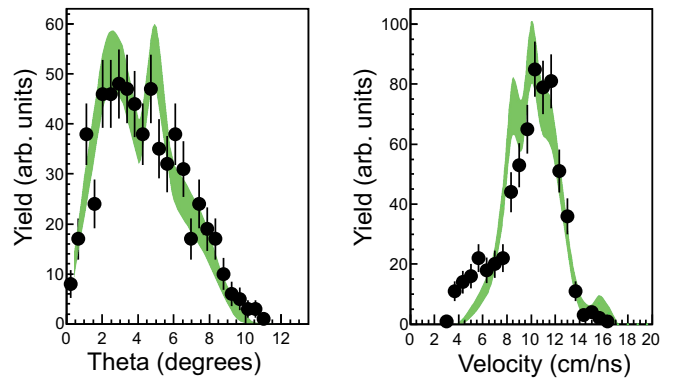}
\caption{\label{f:neut} (Color online) Theta and velocity distributions for the neutron hits within MoNA in coincidence with $Z$~=~4 fragments.  The experimental distributions are shown as the solid circles and the filtered $L$~=~78~MeV CoMD simulations are shown as the filled green contours.}
\end{figure}

Since the fragments measurements did require a neutron hit coincidence trigger, it is important to compare neutron observables between the experiment and simulation to ensure that the fragment yields are not being biased by the trigger condition.  In Fig.~\ref{f:neut} the theta and velocity distributions of the neutron hits from both the experiment and filtered simulation are shown.  The results indicate that the simulation and $\texttt{\sc Geant4}$ filter provide a reasonable description of the neutron distributions and interactions in MoNA.

\begin{figure}
\includegraphics[width=0.45\textwidth]{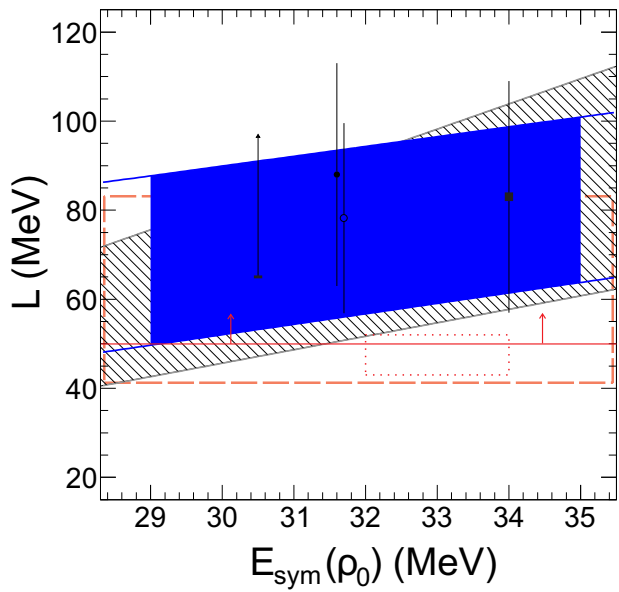}
\caption{\label{f:esym} (Color online) The solid blue area represents the constraints extracted from the present work on the slope ($L$) and magnitude at $\rho_{0}$ of the symmetry energy.  The grey hashed region indicates the heavy-ion collisions constraints from Ref.~\cite{TSANG09}.  Additional constraints from heavy-ion collisions are shown from Refs.~\cite{KOHLEYIMF}, \cite{LI08}, \cite{Shet07}, and~\cite{RUSSOTTO11} as a lower-limit, solid circle, open circle and solid square, respectively.  The $E_{sym}(\rho_{0})$ of the open circle was offset by 0.1~MeV for clarity.  The dotted red line box~\cite{Stei12}, long-dashed red line box~\cite{Steiner13_apj}, and solid red line lower-limit~\cite{Sot12} represent recent constraints extracted from neutron star properties.}
\end{figure}

\begin{figure*}
\includegraphics[width=0.79\textwidth]{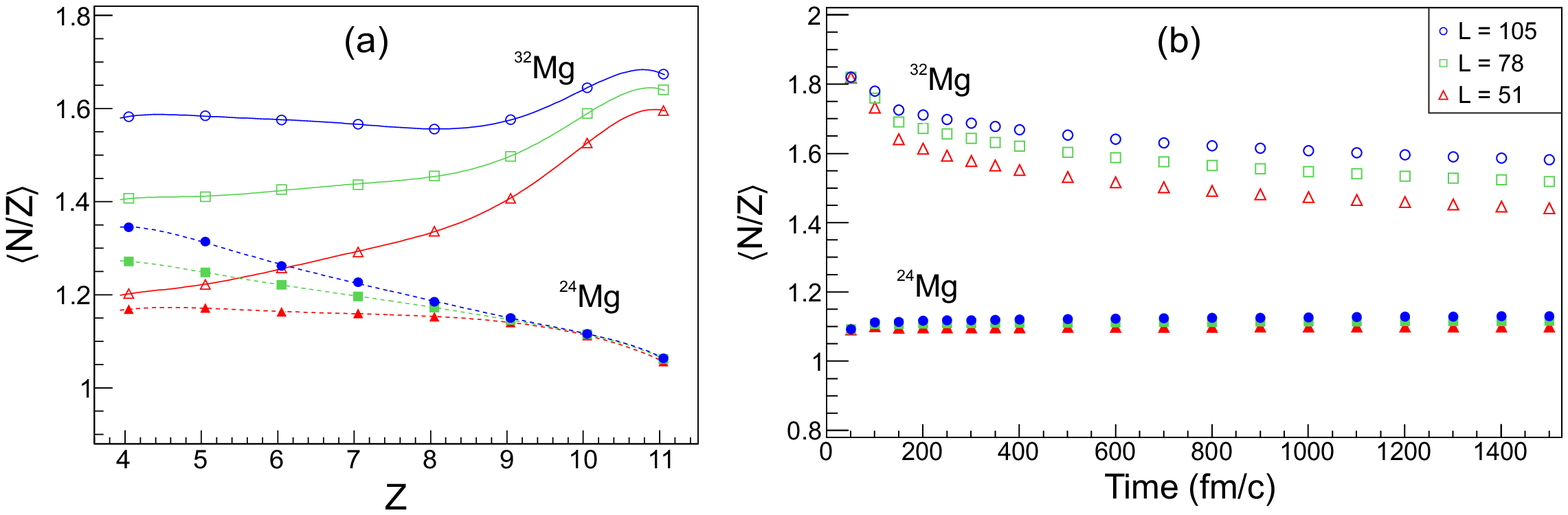}
\caption{\label{f:sim} (Color online) (a) Average $N/Z$ as a function of the charge of the residue fragment.  (b) Average $N/Z$ of $Z$~=~4-11 fragments as a function of the collision time.  The results are shown from the $^{32}$Mg~+~$^{9}$Be and $^{24}$Mg~+~$^{9}$Be CoMD simulations for different slopes of the symmetry energy. }
\end{figure*}

\par
Constraints on the density dependence of the symmetry energy were extracted from the experimental data from comparison with the CoMD simulations.  For both the $\langle N/Z \rangle$ and $Z$ yield plots, the optimum $L$ for reproducing each experimental data point for the $Z$~=~4-9 fragments was determined from the interpolation of the three CoMD simulations ($L = 51, 78, 105$~MeV).  The $Z$~=~10 and 11 fragments were excluded from the analysis to minimize effects due to quasi-elastic and direct reactions which are not driven by the nuclear equation of state.   For example, the experimental $\langle N/Z \rangle$ for $Z = 4$ is 1.67, which is in between the $L = 51$~MeV and 78~MeV CoMD simulations. From the interpolation of the simulation, the experimental data should be best reproduced with $L = 71.3$~MeV.  The procedure of interpolating the optimum $L$ was validated by checking that both the $\langle N/Z \rangle$ and $Z$ yield exhibited a linear dependence with respect to $L$.

\par
After calculating the optimum $L$ for each experimental data point, the average $L$ was determined along with the standard deviation of the mean.  This resulted in $L = 76 \pm 11$~MeV with $E_{sym}(\rho_{0}) = 32$~MeV, where the error is the 2$\sigma$ limit following the standard set by Tsang \textit{et al.}~\cite{TSANG09}.  Additional CoMD calculations were also completed with $E_{sym}(\rho_{0})$~=~29 and 35~MeV with $L$~=~51, 78, and 105~MeV.  The same procedure discussed above was followed for each $E_{sym}(\rho_{0})$ and the average $L$ was calculated.  The 2-dimensional $L$ versus $E_{sym}(\rho_{0})$ constraint from the comparison of the CoMD simulation with the RIB $^{32}$Mg~+~$^{9}$Be experiment is presented in Fig.~\ref{f:esym} and compared with previous heavy-ion collision constraints on the symmetry energy.  The uncertainty of the constraints presented in Fig.~\ref{f:esym} include the statistical $2 \sigma$ error along with a 15$\%$ systematic error estimated from the uncertainty in the yields of fragments with non-central track trajectories through the Sweeper magnet.

The constraints extracted on the density dependence of the symmetry energy are in good agreement with the previous results from heavy-ion collisions as shown in Fig.~\ref{f:esym}.  A consistency is observed indicating a preference for $L \approx 75$~MeV from the reaction studies.  For comparison, recent constraints extracted from neutron star observables are also presented in Fig.~\ref{f:esym}.  The constraints from Sotani~\emph{et al.} (solid red line lower-limit)~\cite{Sot12} and Steiner, Lattimer, and Brown ($2\sigma$ limit, long-dashed red line box)~\cite{Steiner13_apj} agree well with the present work.  The $1\sigma$ limit constraints of Steiner and Gandofli (dotted red line box)~\cite{Stei12} indicate a softer density dependence of the symmetry energy then the heavy-ion collisions constraints.  However, it has been suggested that this constraint was limited by the use of a fixed parameterization for the EoS and may not have accounted for all the systematic uncertainties of the neutron star mass and radius observations~\cite{Steiner13_apj}.  Other observables, such as nuclear binding energies~\cite{Moller12}, isobaric analog states~\cite{Dan09}, neutron skin thicknesses~\cite{Chen10,Abr12}, and pygmy dipole resonances~\cite{Car10}, will be important for providing additional clarity in determining the true form of the density dependence of the symmetry energy~\cite{Tsang12}.  It may also be important to compare the actual forms of the density dependence of the symmetry energy and higher order terms (such as $K_{sym}$) since $L$ is only a representation of the slope at saturation density.

\par
The use of a neutron-rich radioactive ion beam was a critical component in allowing for the constraints on the density dependence of the symmetry energy to be determined.  The $^{32}$Mg projectile has a large $\delta = 0.667$ (for comparison $^{132}$Sn has $\delta = 0.64$) which greatly enhances the sensitivity of the experiment to the symmetry energy as it has a $\delta^{2}$ dependence.  To demonstrate the advantage and importance of using RIBs, the stable beam reaction $^{24}$Mg~+~$^{9}$Be was also simulated using the CoMD model and compared with the $^{32}$Mg~+~$^{9}$Be RIB reaction.  In Fig.~\ref{f:sim}(a) the unfiltered $\langle N/Z \rangle$ as a function of $Z$ for the heavy residue is shown from both the RIB and stable beam reactions.  A drastic increase in the sensitivity to the slope of the symmetry energy ($L$) is observed for the $^{32}$Mg system relative to the $^{24}$Mg system.  The sensitivity of the $\langle N/Z \rangle$ observable to the form of the symmetry energy, measured as the relative difference between the $L = 108$ and 51~MeV simulations, was increased by a factor of 3 by using the RIB in comparison to the stable beam.  The evolution of the $\langle N/Z \rangle$ of the $Z$~=~4 - 11 fragments as a function of time (Fig.~\ref{f:sim}(b)) shows the same increased sensitivity of the the radioactive beam induced reaction to $L$ relative to the stable beam reaction. The evolution of the $^{32}$Mg simulation quickly diverges with respect to $L$ during the initial $\sim 300$~fm/c of the reaction.  From $300-1500$~fm/c additional particle emission occurs from the hot fragments and by $1500$~fm/c the simulation appears to be relatively stable with little change in the $\langle N/Z \rangle$.

\par
\emph{Conclusion.} In summary, the neutron multiplicity and $\langle N/Z \rangle$ of the fragments from the $^{32}$Mg~+~$^{9}$Be were measured using the MoNA~+~Sweeper magnet experimental setup.  Through comparison of the experimental data with the CoMD model constraints on the form of the density dependence of the symmetry energy at subsaturation densities were extracted and are presented in Fig.~\ref{f:esym}.  These results confirm, for the first time, the constraints extracted from Sn~+~Sn heavy-ion collisions by Tsang \emph{et al.}~\cite{TSANG09}.
\par
The comparison of the CoMD simulations for the radioactive $^{32}$Mg and stable $^{24}$Mg induced reactions demonstrated the substantial advantage that neutron-rich RIBs offer experimentalists for probing the density dependence of the symmetry energy. The development of RIB induced experiments will be critical for extracting strict constraints on the form of the symmetry energy at both subsaturation and suprasaturation densities.  The next generation of RIB facilities, such as the Facility for Rare Isotope Beams (FRIB), will greatly improve the opportunities for these RIB induced reactions to realized.

\par
\emph{Acknowledgments.} The authors are grateful to S.~J. Yennello, A.~B. McIntosh, and A. Bonasera for insightful discussions and careful reading of the manuscript.  The authors would like to thank M. Papa and A. Bonasera for use of the CoMD code.  The authors gratefully acknowledge the support of the NSCL operations staff for providing a high quality beam.  This work is supported by the National Science Foundation under Grant Nos. PHY-1102511, PHY-0969173, PHY-0969058, and PHY-1205357.

\par
\emph{Appendix.} The constraints extracted from the RIB induced reactions of this work can be reproduced from the following two linear equations,
\begin{eqnarray*}
    L&=&2.2E_{sym}(\rho_{0}) + 23.9 \\
    L&=&2.35E_{sym}(\rho_{0}) - 18.48
\end{eqnarray*}
which represent the $2\sigma$ limits with the systematic uncertainties included.


%

\end{document}